\documentclass[iop,apj]{emulateapj}
\usepackage[colorlinks,linkcolor={blue},citecolor={blue}]{hyperref}
\usepackage{xcolor}
\usepackage{amsmath}
\usepackage{bm}

\begin{document}

\title{Constraining the inclination of binary mergers from gravitational-wave
 observations}

\author{S.~A.~Usman, J.~C.~Mills, S.~Fairhurst}
\affil{School of Physics and Astronomy, Cardiff University, The Parade, Cardiff,
CF24 3AA, UK}

\begin{abstract}
Much of the information we hope to extract from the gravitational-waves
signatures of compact binaries is only obtainable when we can accurately
constrain the inclination of the source.  In this paper, we discuss in detail a degeneracy
between the measurement of the binary distance and inclination which 
limits our ability to accurately measure the inclination using gravitational 
waves alone.  This degeneracy is exacerbated by the expected distribution
of events in the universe, which leads us to prefer face-on systems at
a greater distance.  We use a simplified model that only considers the binary
distance and orientation, and show that this gives comparable results
to the full parameter estimates obtained from the binary neutron star merger
GW170817.  For the advanced LIGO-Virgo network, it is only signals which are
close to edge-on, with an inclination greater than $\sim 75^{\circ}$ that will be
distinguishable from face-on systems.
For extended networks which have good sensitivity to both gravitational
wave polarizations, for face-on systems we will only be able to constrain 
the inclination of a signal with SNR 20 to be $45^{\circ}$ or less, and even
for loud signals, with SNR of 100, the inclination of a face-on signal
will only be constrained to $30^{\circ}$.  For black hole mergers
observed at cosmological distances, in the absence of higher modes or
orbital precession, the strong degeneracy between inclination and 
distance dominates the uncertainty in measurement of redshift and hence
the masses of the black holes.
\end{abstract}

\maketitle

\newcommand{\samcomment}[1]{\textcolor{violet}{[#1]}}
\newcommand{\camcomment}[1]{\textcolor{green}{[#1]}}
\newcommand{\steve}[1]{\textcolor{blue}{[#1]}}
\newcommand{\checkme}[1]{\textcolor{red}{#1}}


\section{Introduction}
With its ground-breaking detections in the first years of its operation, the
upgraded Laser Interferometer Gravitational-Wave Observatory (LIGO) and Virgo
detectors have opened up the door to discovering new information about the
universe. The collaboration's many gravitational-wave (\emph{GW}) detections
from binary systems, including GW150914 \citep{Abbott:2016blz} and GW170817
\citep{Abbott:2017ntl} have allowed us to draw new insights from these
astrophysical sources. These developments include constraining the nuclear
equation of state \citep{De:2018uhw} and constraining binary black hole
populations \citep{Farr:2017uvj, Roulet:2018jbe, Tiwari:2018qch}. With more detections, we hope
to learn even more about our universe, such as more accurately measuring the
Hubble constant $H_0$ as suggested in \cite{Schutz:1986gp} or
detailing the opening angle for gamma ray bursts (\emph{GRBs}) from binary
neutron star systems (\emph{BNS}) \citep{Clark:2014jpa, Metzger:2017wot,
goldstein_ordinary_2017}.  However, both of these measurements rely on the
accurate measurement of the distance to the binaries and the inclination of their
orbital angular momentum with respect to the line of sight.  A degeneracy
exists between distance and inclination making the measurement of these two
parameters very difficult.  Of the compact binary detections made by LIGO and
Virgo, only the BNS merger GW170817 has had a tightly constrained inclination
and distance. The detection of a kilonova afterglow allowed for an accurate
distance measurement \citep{tanvir_emergence_2017, villar_complete_2017}, breaking the
degeneracy with inclination. When this type of external information is
unavailable, the degeneracy severely limits our ability to measure these
parameters.

In this paper, we will show that this degeneracy is typical for binary mergers.
The measured amplitude and phase of the gravitational-wave signal encode the
properties of the binary.  In particular, it is the differing amplitude of the
two polarizations of the gravitational waveform that allow us to determine the
binary inclination. However, the plus ($+$) and cross ($\times$) polarizations
have nearly identical amplitudes at small inclination angles (less than
45$^{\circ}$) and significantly lower amplitudes at large inclination angles
(greater than 45$^{\circ}$). This leads to two simple observations: first, that
the signal is strongest for binaries which are close to face-on ($\iota \sim
0^{\circ}$) or face-away ($\iota \sim 180^{\circ}$) and thus we will be observationally
biased to detecting binaries whose orbital angular momentum is well-aligned (or
anti-aligned) with the line of sight \citep{nissanke_exploring_2010,
schutz_networks_2011}. Second, for small angles, the amplitudes of the two
polarizations are close to equal and we cannot measure distance or inclination
separately.  Therefore, for the majority of detections, this face-on degeneracy
will limit our ability to constrain both electromagnetic (\emph{EM}) emission
models and the Hubble constant. There are various ways to break this
degeneracy, such as using the EM measured distance or using jet modelling to
constrain the opening angle.  These techniques were used to improve the
constraints on the inclination and distance for the BNS merger GW170817
\citep{cantiello_precise_2018, mandel_orbit_2018, finstad_measuring_2018,
Abbott:2018wiz, guidorzi_improved_2017}.

Since an inclined binary system would produce both a high-amplitude plus
polarization and a lower-amplitude cross polarization, creating a network of
detectors which is sensitive to both the plus and cross polarization has been
suggested to constrain the inclination using only gravitational waves
\citep{blair_science_2008}. A single detector is sensitive to just one
polarization. Hanford and Livingston are almost aligned, and see essentially
the same polarization, while Virgo is anti-aligned and is sensitive to the
orthogonal polarization. The addition of Kagra \citep{aso:2013eba} and India
\citep{Reitze:2011} would further increase the network's sensitivity to the
orthogonal polarization. Thus it is hoped this network could better constrain
the inclination angle and distance. We examine this possibility of constraining
the inclination using only the measurement of the two GW polarizations.

There have been many studies looking at inclination constraints. From the GRB
perspective they are largely divided into two groups: the first focuses on
exploring the possibility of nailing down the viewing angle by comparing the
rate of GRB sources observed in GWs with those in gamma rays
\citep{williams_constraints_2017, Clark:2014jpa, Williamson:2014wma}.  The
second focuses on measurements for individual detections, mainly in the case
where the event has been three dimensionally localized by an EM counterpart
\citep{seto_prospects_2007, arun_synergy_2014}. 
In  \citep{chen_measuring_2018} it was observed that the inclination measurement 
is poor for binaries with an inclination less than seventy degrees when there is no
redshift information. They attribute this to a combination of the degeneracy between 
distance and inclination, and the prior on the distance. Here we explore the origin of 
the degeneracy in detail and discuss the importance of an additional degeneracy
when the binary is circularly polarized \cite{Fairhurst:2017mvj}.

Inclination
constraints have also been discussed in the context of distance estimates for
cosmology \citep{markovic_possibility_1993, nissanke_exploring_2010,
chen_gamma-ray-burst_2013} and as part of wider parameter estimation
investigations \citep{cutler_gravitational_1994, veitch_estimating_2012}. It
was noted in \cite{nissanke_exploring_2010} that adding detectors to a network
did not seem to greatly improve the inclination measurement. Here we push this
to this extreme by including all current and proposed future ground-based
observatories.  In particular, we investigate a network that would measure both
polarizations equally as would be expected over the majority of the sky for the
Einstein Telescope (\emph{ET}) \citep{punturo_third_2010}.

\section{Measuring Distance and Inclination}

When a gravitational-wave signal is observed in the data from the LIGO and
Virgo instruments, the goal is to obtain estimates for the parameters that
describe the waveform.  Typically, Bayesian inference \citep{Veitch:2014wba,
Christensen:2004bc, Rover:2006ni} is used to obtain a posterior distribution
for the parameters of the system $\bm{\theta}$ given the observed data $\mathbf{d}$.  As
described in detail in \cite{Maggiore:2008gw}, the likelihood of obtaining data
$\mathbf{d}$ given the presence of a signal $h(\theta)$, and under the
assumption of Gaussian noise characterized by a power spectrum $S(f)$, is
\begin{equation}
\label{eq:like}
\Lambda(\mathbf{d} | \bm{\theta}) \propto \exp\left[ - \frac{1}{2} \left(\mathbf{d} - h(\bm{\theta})| \mathbf{d} - h(\bm{\theta}) \right)\right] \, .  
\end{equation}
Here, we have introduced the weighted inner product
\begin{equation}
\label{eq:m_filter}
\left( a| b\right) := 4 \mathrm{Re} \int_{0}^{f_{\mathrm{max}}} \frac{\tilde{a}(f)\tilde{b}(f)^{\star}}{S(f)} \; \text{d}f \, . 
\end{equation}

The likelihood for a network of detectors is simply the product of likelihoods
for the individual detectors:
\begin{equation}
\Lambda(\mathbf{d} | \bm{\theta}) \propto \exp\left[ - \frac{1}{2} \sum_{i \in \mathrm{dets}} \left(\mathbf{d}_{i} - h_{i}(\bm{\theta})| \mathbf{d}_{i} - h_{i}(\bm{\theta}) \right)\right] \, .
\end{equation}

The posterior distribution for parameters $\bm{\theta}$ given the data $\mathbf{d}$
is given as
\begin{equation}
p(\bm{\theta} | \mathbf{d}) \propto \Lambda(\mathbf{d} | \bm{\theta} ) p(\bm{\theta} ),
\end{equation}
where $p(\bm{\theta} )$ is the prior distribution for the parameters. The
posterior distributions are typically calculated by performing a stochastic
sampling of the distribution \citep{Christensen:2001cr, Christensen:2004bc,
Rover:2006ni, vanderSluys:2007st, vanderSluys:2008qx}.  Distributions for a
subset of parameters are obtained by marginalizing, or integrating out, the
additional parameters.

In this analysis, we are interested in obtaining the joint distribution of the
luminosity distance $d_L$ and binary inclination $\iota$.  This is calculated
as
\begin{equation}
p(d_L, \cos\iota | \mathbf{d}) = \int d\bm{\mu} \Lambda(\mathbf{d} | \bm{\mu}, d_L, \cos\iota) p(\bm{\mu}, d_L, \iota)
\end{equation}
Typically, $\bm{\mu}$ contains all parameters describing the system, including
the masses, spins, sky location, orientation and parameters describing the
nuclear equation of state. For our work, we consider a simplified model, for
which the only additional parameters $\bm{\mu}$ are the binary's polarization
$\psi$ and coalescence phase $\phi_{o}$.  We choose uniform priors on these
parameters, as well as a uniform prior on $\cos\iota$, which leads to a uniform
distribution of binary orientation.  Furthermore, we use a uniform-in-volume
prior for the distance $p(d_L) \propto d_L^{2}$.  For binaries at greater
distance, we need to take into account cosmological effects and use a prior
with sources uniform in comoving volume and merging at a constant local rate.
At even greater distances, the local merger rate would follow the star
formation rate \citep{madau_cosmic_2014}, which peaks at $z \sim 2$. We take
this into account later in this paper for binary black hole systems,
(\emph{BBH}), which can be detected throughout the universe with future detectors.

In our approximation, we fix the sky location and arrival time of the signal,
as well as the masses and spins of the system.  Fixing the sky location is
reasonable, as one of the main motivations for this work is to investigate the
accuracy of gravitational-wave measurements of distance and inclination after
the signal has already been identified and localized by the detector network.
We also investigate how inclination measurements from gravitational-wave
observations can be combined with electromagnetic observations. An unknown sky
location will only lead to larger uncertainties in the distance and inclination
measurements arising from varying detector sensitivities over the sky.

While the masses and spins of the binary will not be known, in most cases these
parameters have little impact on the inferred distance and inclination. Binary
neutron star systems are in-band in ground-based detectors for a large number
of cycles, $\mathcal{O}(10^5-10^6)$, allowing the accurate measurement of the
phase evolution of the binary.  Hence the chirp mass ${\mathcal {M}}$ --- the
parameter determining the leading order phase evolution --- is measured with
great precision.  For BNS, the GW amplitude scales as ${\mathcal
{M}}^{\frac{5}{6}}$, so uncertainty in mass has no effect on the distance
$d_L$. In the analysis presented here, we focus only on the dominant
gravitational-wave emission at twice the orbital frequency.  For unequal-mass
systems, the other gravitational-wave harmonics can significantly affect the
waveform, particularly when the binary has a high mass ratio, i.e. one of the
compact objects is significantly more massive than the
other~\citep{Capano2014}.  This can lead to improvements in the measurement of
the binary orientation~\citep{london_first_2017}.

Spins which are misaligned with the orbital angular momentum lead to precession
of the binary orbit \citep{Apostolatos1994} which can, in principle, lead to an
improved measurement of the binary orientation.  To date, there is no evidence
for precession in the observed GW signals \citep{TheLIGOScientific:2016pea,
Abbott:2017vtc, Abbott:2017gyy, Abbott:2017oio, TheLIGOScientific:2017qsa}, so the approximations discussed here
would therefore be applicable.  Furthermore, neutron stars are not expected to
achieve a spin high enough to have observable precession.

\begin{figure*}[t]
\includegraphics[width=0.99\linewidth]{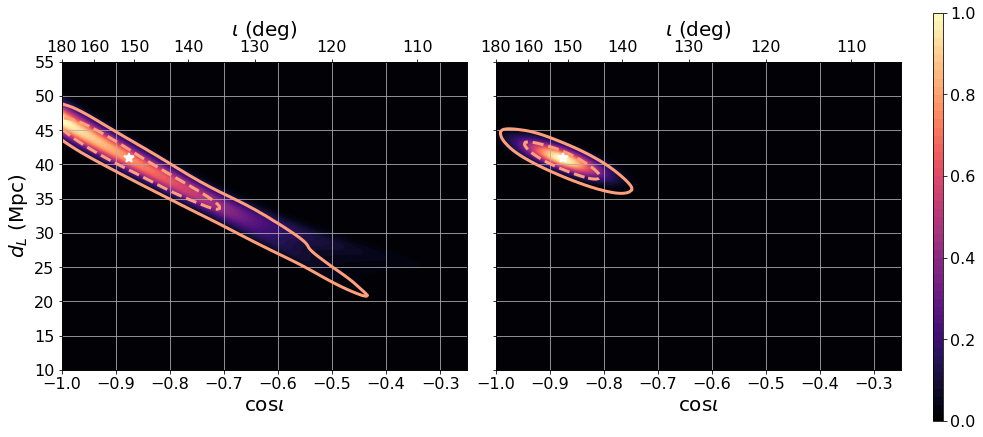}
\caption{The marginalized posterior distribution for the distance and
inclination of the binary neutron star system GW170817, detected with an
alignment factor $\alpha \sim 0.13$ and signal to noise ratio $\rho \sim 32$.
The left plot was generated using only the data from gravitational-wave
detectors, while the right plot also uses the independent distance measurement
($40.7$ Mpc, $\pm 2.4$ Mpc at 90\% confidence) from electromagnetic
observations. The coloured portion of the plot shows the probability
distribution obtained using our approximate analysis, normalized such that the
peak probability is 1. The orange contours represent the 90\% and the 50\%
confidence intervals obtained by performing the full analysis of the LIGO-Virgo
data (posterior samples are publicly available here:
\url{https://dcc.ligo.org/LIGO-P1800061/public})
\citep{Abbott:2018wiz}.}
\label{fig:GW170817}
\end{figure*}

To verify that fixing the masses and spins has limited impact on the recovered
distance and inclination, we compare results from our model with those from the
full parameter estimation of GW170817.  We recreate the posterior distribution
for the multi-messenger signal GW170817, with and without distance information
from the coincident electromagnetic signal, and compare it to the full,
Bayesian parameter estimation, with a fixed sky location, using the observed
LIGO and Virgo data \citep{Abbott:2018wiz}.
The results are shown in Figure~\ref{fig:GW170817}. To generate our results, we
approximate the data $\mathbf{d}$ by a gravitational-wave signal at a distance of $d_L =
40.7$ Mpc \citep{cantiello_precise_2018} and an inclination of $153^{\circ}$
\citep{Abbott:2018wiz}. We then generate a
posterior distribution for the four dimensional parameter space of distance
$d_L$, inclination $\iota$, polarization $\psi$ and coalescence phase $\phi_0$.
From this we calculate the posterior distribution, $p(d_L, \iota | \mathbf{d})$ by
marginalizing over the polarization and phase angles. As is clear from the
figure, our approximate method gives a posterior on distance and inclination
which is in excellent agreement with the full results from the real data.%
\footnote{We note that the results in
\cite{Abbott:2018wiz} show this distribution
as a function of inclination $\iota$ instead of $\cos\iota$.  This leads to a
different distribution, and different 90\% confidence intervals as these are
defined to be the minimum range that contains 90\% of the probability, and this
is dependent upon variable choice.  As we discuss later, there is no evidence
in the GW data alone that the signal is not face-on, and since the prior is
flat in $\cos \iota$ we believe that plotting the distribution against $\cos
\iota$ leads to a clearer understanding of the distribution.}

The results in Figure~\ref{fig:GW170817} show an example of the degeneracy in
the measured values of the distance and binary inclination.  The 50\%
confidence interval includes both a face-away binary at a distance of $45$ Mpc
and a binary inclined at $135^{\circ}$ at a distance of $35$ Mpc.  It is only
when the gravitational-wave data is combined with the electromagnetically
determined distance $45 \pm 2.4$ Mpc \citep{cantiello_precise_2018} that the
binary inclination can be accurately inferred. The degeneracy between distance
and inclination arises directly from the dependence on the gravitational
waveform on these parameters, and has been discussed several times previously
\citep{markovic_possibility_1993, cutler_gravitational_1994,
nissanke_exploring_2010}.

To understand why distance and inclination are degenerate, we must look to the
waveform of gravitational waves emitted from a binary system. The
gravitational-wave signal, $h(t)$, incident on a gravitational-wave detector is
given by \citep{Thorne:1987yg}:
\begin{equation}
h(t) = F_+(\alpha, \delta, \chi) h_+(t) + F_{\times}(\alpha, \delta, \chi) h_{\times}(t),
\label{eq:detector_response}
\end{equation}
where $F_+$ and $F_{\times}$ are the detector response to the plus and cross
polarizations, respectively. The detector responses depend on the location
$(\alpha, \delta)$ of the source.  In addition, we must specify a polarization
angle $\chi$ to fully specify the \textit{radiation frame}.  It is common
\citep{klimenko_constraint_2005, harry_targeted_2011} to define a dominant
polarization frame, for which the detector network is maximally sensitive to
the plus polarization.  With this choice, we can naturally characterize the
network by its overall sensitivity and the relative sensitivity to the second
polarization \citep{klimenko_constraint_2005, mills_localization_2018}.  This
simplifies the comparison of different networks.

\begin{figure}[t]
\includegraphics[width=\linewidth]{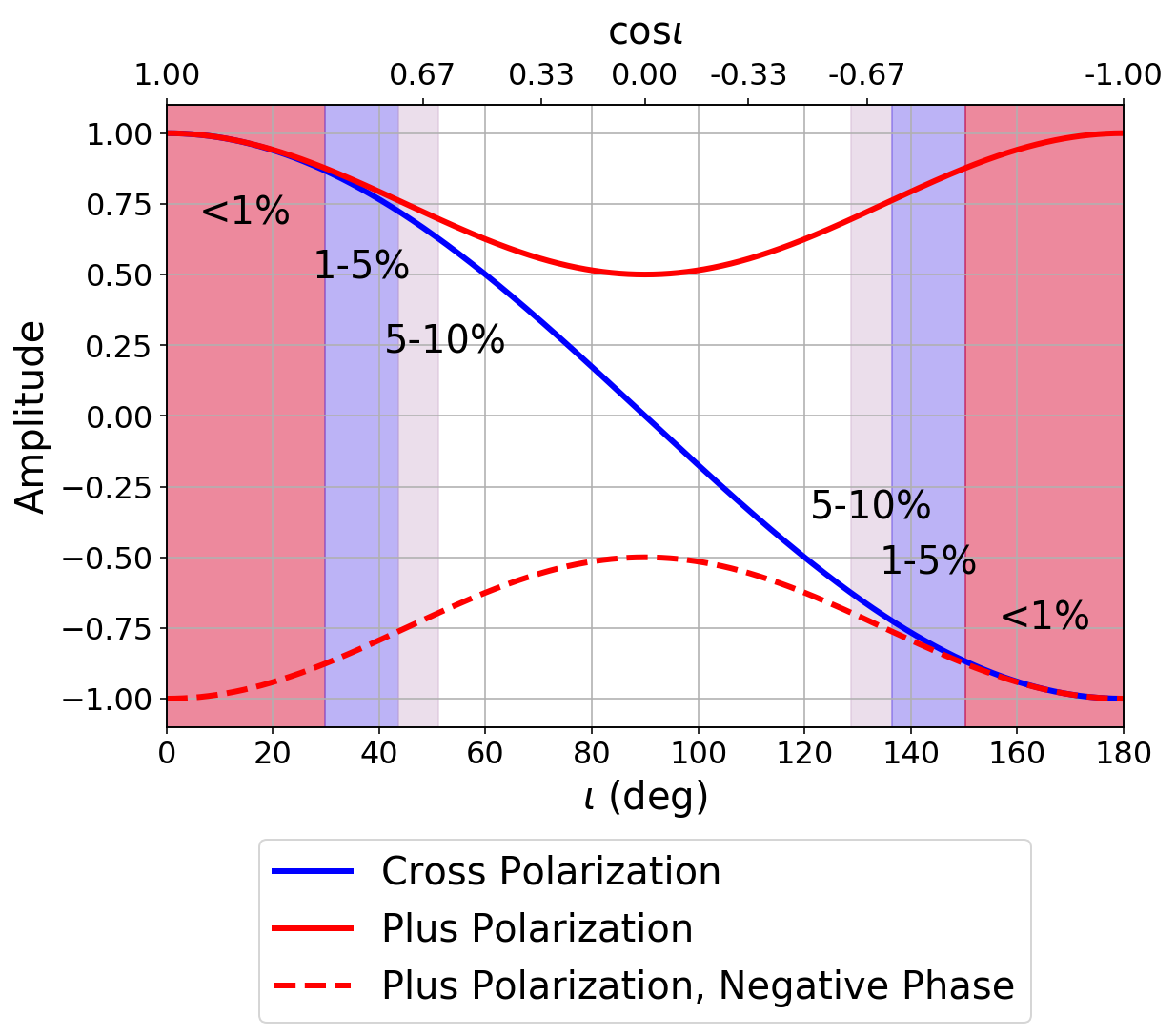}
\caption{The relative contributions of the plus and cross polarizations to a
gravitational-wave signal, dependent on the inclination. The red solid line
indicates the amplitude of the plus polarization, while the dashed red solid
line indicates the amplitude for the plus polarization with a negative phase.
The blue solid line indicates the amplitude of the cross polarization. The
shaded regions show the percent differences between the plus and cross
polarizations. The red portion represents when the plus and cross polarization
are less than 1\% different. The blue region represents where the polarizations
are between 1\% and 5\% different.  The grey region represents where the
polarizations are between 5\% and 10\% different.}
\label{fig:prop}
\end{figure}

For a waveform where it is appropriate to neglect higher order modes and
precession, the two polarizations given in Equation~\ref{eq:detector_response}
can be expressed in terms of the two orthogonal phases of the waveform:
\begin{align}
h_+(t) = \mathcal{A}^1 h_0(t) + \mathcal{A}^3 h_{\frac{\pi}{2}}(t)\\
h_{\times}(t) = \mathcal{A}^2 h_0(t) + \mathcal{A}^4 h_{\frac{\pi}{2}}(t)
\end{align}
where $\widetilde{h}_{\frac{\pi}{2}}(f)=i\widetilde{h}_0(f)$.  The
$\mathcal{A}^i$ are overall amplitude parameters, and depend on the distance
$D$, inclination $\iota$, polarization $\psi$ and coalescence phase $\phi_0$
\citep{Cornish2007,Bose1999}:
\begin{align}
\mathcal{A}^1 &=\mathcal{A}_+\cos 2\phi_0 \cos 2\psi- \mathcal{A}_{\times}\sin 2\phi_0 \sin 2\psi \\
\mathcal{A}^2 &=\mathcal{A}_+\cos 2\phi_0 \sin 2\psi + \mathcal{A}_{\times}\sin 2\phi_0 \cos
2\psi \\
\mathcal{A}^3 &=-\mathcal{A}_+\sin 2\phi_0 \cos 2\psi -\mathcal{A}_{\times}\cos 2\phi_0 \sin 2\psi \\
\mathcal{A}^4 &=-\mathcal{A}_+\sin 2\phi_0 \sin 2\psi +\mathcal{A}_{\times}\cos 2\phi_0 \cos
2\psi,
\end{align}
where $\mathcal{A}_+$ and $\mathcal{A}_{\times}$ are amplitudes for the plus
and cross polarizations in the \emph{source frame}, which is aligned with the
binary's orbital angular momentum.  They are given by:
\begin{gather}
\mathcal{A}_+ = \frac{d_0}{d_L}\frac{1+\cos^2 \iota}{2}\\
\mathcal{A}_{\times} = \frac{d_0}{d_L}\cos \iota,
\end{gather}
where $d_L$ is the luminosity distance and $d_0$ is the reference luminosity
distance.  The variation of the two polarization amplitudes with inclination
$\iota$ is shown in Figure~\ref{fig:prop}. We note that there is an arbitrary
choice of the \textit{radiation frame} and this will affect the value of the
angles $\psi$ and $\chi$ and consequently the values of the $\mathcal{A}^{i}$.
However, the signal observed at the detectors is independent of this choice.

In principle, we should be able to measure all four of the amplitude parameters
by accurately measuring both the amplitude and phase of both the plus and cross
polarizations of a gravitational wave. From here, we could then infer the
distance and orientation of the source binary. However, degeneracies in
parameters limits our ability to accurately measure these parameters.

In order to identify the inclination of the binary system using the
polarizations of the gravitational wave, we must distinguish the contributions
of the plus and cross polarizations. When the binary system is near face-on or
face-away, the two amplitudes $\mathcal{A}_{+}$ and $\mathcal{A}_{\times}$ have
nearly identical contributions to the overall gravitational-wave amplitude. In
Figure~\ref{fig:prop}, we see the relative difference between plus and cross is
less than $1\%$ for inclinations less than $30^{\circ}$ (or greater than
$150^{\circ}$) and $5\%$ for inclinations less than $45^{\circ}$ (or greater
than $135^{\circ}$). This is the main factor that leads to the strong
degeneracy in the measurement of the distance and inclination.

As we have already described, gravitational-wave detectors with limited
sensitivity will preferentially observe signals which are close to face-on or
face-off. In addition, when the binary is close to face-on and the emission is
circularly polarized, the waveform is described by a single overall amplitude
and phase (as the two polarizations are equal, up to a phase difference of
$\pm90^{\circ}$).  Thus it is no longer possible to measure both the
polarization $\psi$ and phase at coalescence $\phi_0$ of the binary, but only
the combination $\phi_0 \pm \psi$ (with the $+/-$ for face-on/away binaries
respectively).  This degeneracy, combined with the distance prior, leads to a
significantly larger volume of parameter-space which is consistent with
face-on, rather than edge-on systems.

\begin{figure*}[t]
\includegraphics[width=0.5\linewidth]{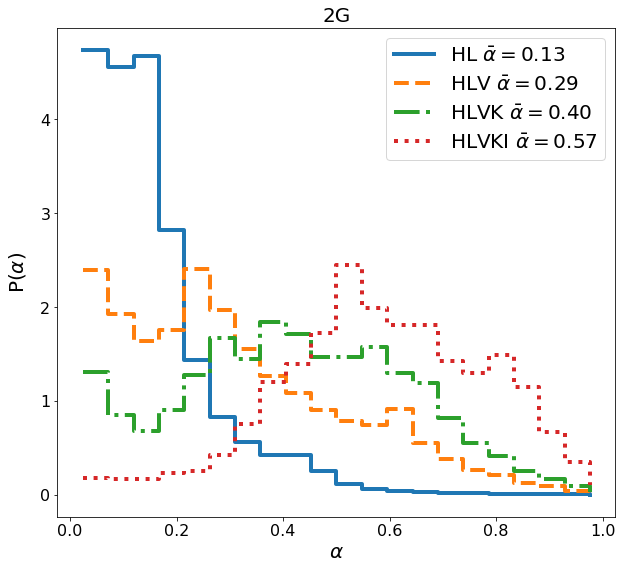}
\includegraphics[width=0.5\linewidth]{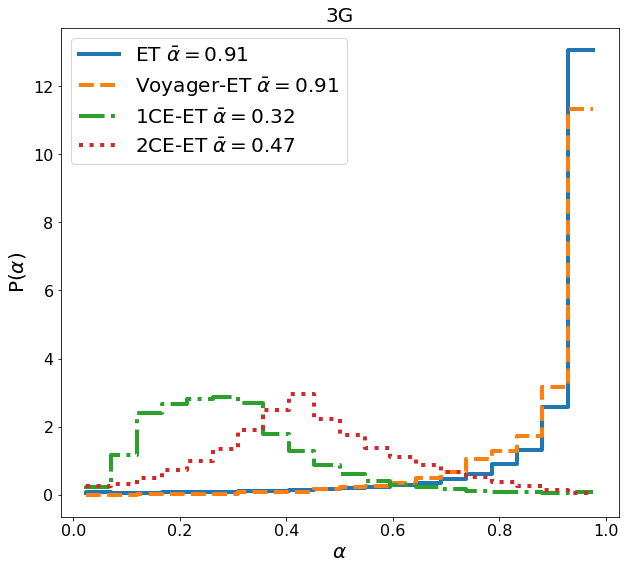}
\caption{The relative sensitivity of detector networks to the second
polarization, as encoded in the parameter $\alpha$, defined through $F_{\times}
= \alpha F_{+}$ (in the dominant polarization frame where the network is
maximally sensitive to the plus polarization). The left plot shows the expected
distribution of $\alpha$ for second-generation gravitational-wave networks,
while the right plot shows the distribution for potential third generation
networks. In both cases, the distribution is the expected distribution for a
population of events, distributed uniformly in volume, and observed above
threshold in the detector network. Thus, directions of good network sensitivity
are more highly weighted.  The second generation networks considered are LIGO
Hanford and Livingston (HL); two LIGO detectors and Virgo (HLV); LIGO-Virgo and
KAGRA (HLVK) and LIGO-Virgo-KAGRA with LIGO-India (HLVKI). As more detectors are
added to the network, the average sensitivity to the second polarization
increases. The right plot shows results for the Einstein Telescope (ET), which
is comprised of three 60-degree interferometers, ET and three LIGO-Voyager
detectors (Voyager-ET) and ET with either one or two Cosmic Explorer detectors
(1CE-ET and 2CE-ET).  As the ET detector has good sensitivity to both
polarizations, networks where ET is the most sensitive detector will have large
values of $\alpha$. Third generation target noise curves are taken from
\cite{Evans:2016mbw}.} \label{fig:alphas} \end{figure*}

To exclude face-on binaries from a marginalized posterior probability
distribution on the inclination, the network must accurately measure the
amplitude and phase of both of the polarizations. In general,
gravitational-wave detectors are not equally sensitive to the two
polarizations.  For a given sky location, we can define the plus polarization
as the linear combination we are most sensitive to and then calculate the
relative sensitivity of $\times$.  We can think of this as a detector network
comprised of a long plus-detector and a shorter cross-detector (a factor of
$\alpha$ shorter). Thus we can estimate the proportional sensitivity to the
second polarization, called the network alignment factor
\citep{klimenko_constraint_2005}, through the relation $F_{\times} = \alpha
F_+$, where $\alpha$ varies between 0 and 1.  Therefore the sensitivity of the
network to the second polarization can be determined by looking at the values
of $\alpha$ over the sky.

Figure~\ref{fig:alphas} shows the distribution of alphas for various detector
networks. As might be expected, the sensitivity to the second polarization
increases as more detectors are added to the network.  For the two LIGO
detectors, the typical value is $\alpha \sim 0.1$ because the two detectors
have very similar orientations.  When the Virgo detector is added to the
network, the mode is $\alpha \sim 0.3$ and this increases to $\alpha \sim 0.5$
when KAGRA and LIGO India join the network.  The Einstein telescope is a
proposed future detector with a triangular configuration
\citep{punturo_third_2010}.  For an overhead source, ET is equally sensitive to
both polarizations, giving $\alpha = 1$.  While ET does not have equal
sensitivity to both polarizations over the whole sky, the majority of signals
will be observed with $\alpha > 0.9$.  For the future networks, we consider an
ET detector complemented by either the advanced LIGO detectors with sensitivity
improved by around a factor of three (LIGO Voyager), or by one or two Cosmic
Explorer detectors \citep{Evans:2016mbw, mills_localization_2018}.  When the ET
detector dominates the network's sensitivity, we have excellent measurement of
both polarizations but, in the CE-ET networks where CE is more sensitive, the
sensitivity to the second polarization is comparable to the current networks.

\section{Accuracy of measuring distance and inclination}

Now that we understand how the degeneracy between inclination and distance
arises, we can explore the expected accuracy with which these parameters will
be measured in various gravitational-wave detector networks.  For concreteness,
in the examples that follow, we fix the SNR of the signals to be 12.  While
this might seem low, we note that for a detection threshold of 8, the
\textit{mean} SNR observed from a uniform-in-volume population would be 12
\citep{schutz_networks_2011}.  We discuss higher SNR signals later in the
paper.  Rather than specifying a network and sky location, we instead
investigate the ability to measure distance and inclination as we vary the
network's relative sensitivity to the second polarization, encoded in the
variable $\alpha$.  For convenience, we fix the masses of the system to be $1.4
M_{\odot}$ and set the sensitivity of detector network to the plus polarization
of GW to be equal to that of a single advanced LIGO detector at design
sensitivity for an overhead source.  This places a face-on system at
approximately 300 Mpc at SNR of 12.  For inclined systems, the distance will be
smaller to ensure that the network still receives an SNR of 12.  While we have
fixed the masses and detector sensitivities to make the plots, the results are
essentially independent of these choices, up to an overall rescaling of the
distance.  Thus the results will be applicable to any system for which it is
reasonable to neglect precession effects and the impact of higher modes in the
gravitational waveform.

\begin{figure*}[t]
\includegraphics[width=\linewidth]{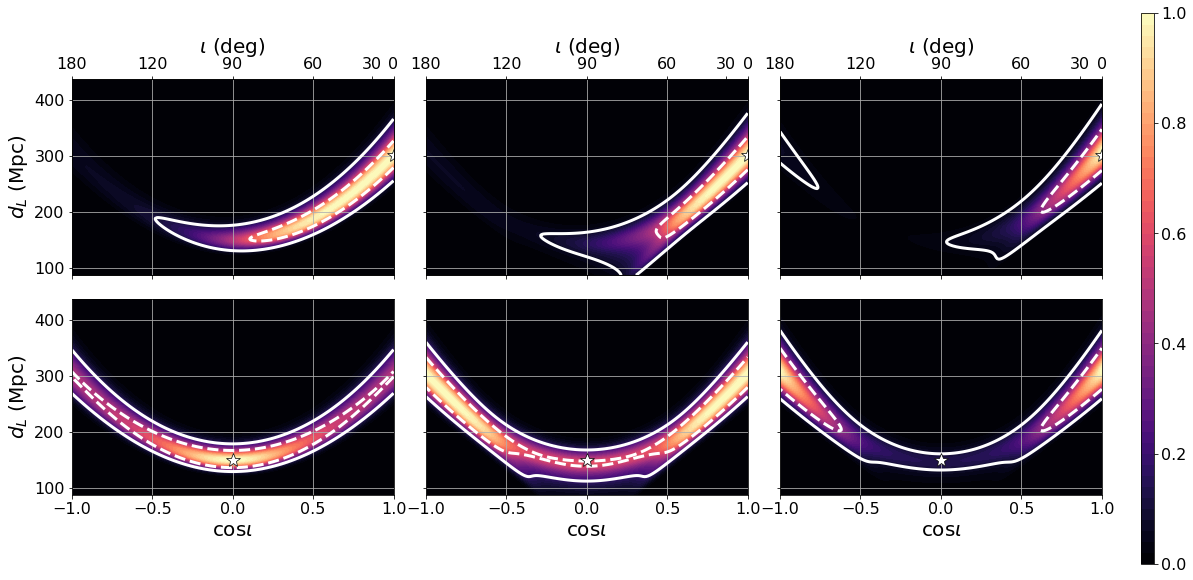}
\caption{The progression of the probability distributions over a $\cos\iota$
and distance parameter space for a signal detected with alignment factor
$\alpha = 0.1$ and signal to noise ratio $\rho = 12$. The top panel shows the
distribution for an face-on signal. The bottom panel shows the distribution for
an edge-on signal. The leftmost plots are the distribution for \emph{only} the
likelihood.  This is generated by calculating the SNR fall-off over the
parameter space.  Since we have not yet marginalized over the phase $\phi$ and
polarization $\psi$, the orientation angles are set to zero here. The middle
plots show how these distribution change when marginalizing over $\psi$ and
$\phi$. Lastly, the rightmost plots are the complete probability distribution,
calculated by applying a distance-squared weighting to the likelihood. This is
to account for the expectation that binary systems are distributed uniformly in
volume. Recall that $\alpha = 0.1$ is the mode sensitivity for the
Hanford-Livingston network.  The white star represents the hypothetical signal.
The white contours represent the 50\% and 90\% confidence intervals obtained
from our simplified model. Note that these contours do not represent the
results of full parameter estimation, as they did in Figure \ref{fig:GW170817}.
From these plots, we can see that at this $\alpha$, a side-on signal is
indistinguishable from a face-on/face-away signal.}
\label{fig:alpha0.1}
\end{figure*}

Let us begin by considering a network with relatively poor sensitivity to the
second GW polarization, with $F_{\times} = 0.1 F_{+} $.  This is typical for
the LIGO Hanford-Livingston network, and is common for the LIGO-Virgo network,
as described in Figure~\ref{fig:alphas} We consider two signals, both with SNR
of 12, but one which is face-on ($\iota = 0$) at a distance of $300$Mpc while
the second is edge-on ($\iota = 90^{\circ}$) at a distance of $150$ Mpc and a
polarization angle of $\psi = 0$ so that the GW power is contained in the plus
polarization.  The first column of figures in Figure~\ref{fig:alpha0.1} shows
the likelihood, maximized over $\phi_0$ and $\psi$, across the
distance-inclination plane.  Note that the contours here are calculated for our
simplified model and do not represent the results of full parameter estimation
analyses, as they did in Figure~\ref{fig:GW170817}. As expected, the maximum
likelihood occurs at values of distance and inclination which exactly match the
signal.  We observe a degeneracy in distance and inclination, so that there is
some support for the edge-on binary to be face-on (or face-away).  There is
also degeneracy for the face-on binary, which is marginally consistent with an
edge-on binary, but face-away orientation can be excluded.  With an SNR of 12
and $\alpha=0.1$, for a face-on signal we expect an SNR of about $1.2$ in the
cross polarization.  These results show that the presence or absence of this
signal is sufficient to down-weight, but not exclude, an edge-on orientation
when the source is really face-on, and vice-versa.  For a face-away system, the
expected signal in the cross polarization is the same amplitude, but entirely
out of phase from the face-on system, and this is sufficient to distinguish the
two.

In the second column, we show the likelihood, marginalized over the
polarization and phase angles.  This marginalization does not have a
significant impact on the face-on binary, but completely changes the
distribution for the edge-on binary --- with the marginal likelihood now peaked
at $\cos\iota = \pm 1$.  Typically, we would expect to be able to measure the
two phase angles with accuracy $\sim 1/\rho$ thus to a crude approximation,
marginalizing over the phase angles would give a contribution $\approx
(1/\rho^{2}) \Lambda_{\mathrm{max}}$, where $\Lambda_{\mathrm{max}}$ is the
maximum likelihood.  When the binary is recovered (nearly) face-on the two
amplitudes $\mathcal{A}_{+, \times}$ are (nearly) equal.  Consequently, the
signal is circularly polarized, with the phase determined by $\phi_0 + \psi$.
Changing the value of $\phi_0 - \psi$ has no effect on the waveform.  Thus,
when marginalizing over the polarization and phase, we obtain a factor $\sim
(\pi / \rho) \Lambda_{\mathrm{max}}$.  Thus, for this signal at SNR 12,
marginalizing of the polarization and phase will lead to a relative increase of
nearly 40 in favour of the face-on signal.

Finally, in the third column, we include the distance prior by re-weighting by
$d_L^{2}$ to place sources uniformly in volume.  This gives an additional
factor of four weighting in favour of the face-on signal over the edge-on one.
Once all these weightings are taken into account, the probability distributions
between a face-on and edge-on signal are similar for a network with this
sensitivity.  The edge-on signal has slightly more support at $\cos \iota
\approx 0$, and this is still included at $90\%$ confidence.  Additionally, the
edge-on signal is consistent with either a face-on or face-away orientation.
It may seem strange that we will not recover the parameters of the edge-on
system accurately.  However, this is appropriate.  As we have discussed, the
volume of parameter space consistent with a face-on system is significantly
larger than for the edge-on case.  Thus, even if we observe a signal that is
entirely consistent with an edge-on system, it is more likely that this is due
to a face-on system and noise fluctuations leading to the observed signal than
it is that the signal is coming from an edge-on system. 

\begin{figure*}[t]
\includegraphics[width=\linewidth]{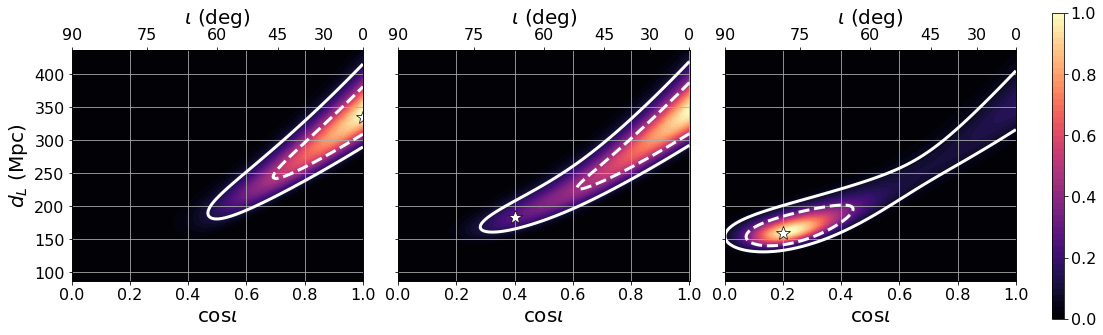}
\caption{The probability distribution over a $\cos\iota$ and distance parameter
space for a signal detected with alignment factor $\alpha = 0.5$ and signal to
noise ratio $\rho = 12$.  The white star represents the injected signal. The
white contours represent the 50\% and 90\% confidence intervals obtained from
our simplified model. Note that these contours do not represent the results of
full parameter estimation, as they did in Figure \ref{fig:GW170817}. A face-on
signal (where $\cos\iota = 1$) returns a nearly identical probability
distribution of the parameter space as a signal from a binary with an
inclination of about 66 degrees ($\cos\iota = 0.4$). For inclinations in the
range $0.1 < \cos\iota < 0.4$, though the distribution now peaks at the correct
inclination, there is support extending across from face-on to an inclination
of  $\iota \sim 80^{\circ}-90^{\circ}$. In these cases it is not possible to
distinguish the binary inclination. The signal is only clearly identified as
\emph{not} face-on after $\cos\iota<0.1$.}
\label{fig:alpha0.5}
\end{figure*}

Our next example investigates differing inclinations for a signal detected by a
network with an $F_{\times} = 0.5 F_{+}$, a network with half the sensitivity
to the cross polarization as the plus polarization.  This is the predicted mean
sensitivity expected for the best near-future detector network consisting of
the Hanford, Livingston, Virgo, KAGRA and LIGO-India detectors. Again, the SNR
is set to 12 for all hypothetical signals, and now we consider three different
inclinations: $\iota = 0$ (face-on) and two inclined signals, one with
$\iota=66^{\circ}$ and the other with $\iota = 78^{\circ}$.  In Figure
\ref{fig:alpha0.5}, we show the posterior distribution for distance and
inclination for the three cases.  Here, we have marginalized over the phase
angles and included the distance prior weighting, so the plots are equivalent
to the third column of plots in Figure \ref{fig:alpha0.1}.

The leftmost plot shows the probability distribution for a face-on signal. This
distribution is similar to the one for $\alpha = 0.1$, though now the most
inclined and face-away points in parameter space are excluded from the 90\%
credible region.  The second plot is for a binary inclined at $66^{\circ}$
($\cos\iota=0.4$).  Here, the peak of the inclination distribution corresponds
to a face-on system and, indeed, the posterior is nearly identical to that
obtained for the face-on system.  Thus, for a typical system with
close-to-threshold SNR we will remain unable to distinguish between face-on
signals and those inclined at $60^{\circ}$ based on gravitational-wave
observations alone.  The best near-future detector therefore would be unable to
measure a difference in inclination between these two hypothetical signals.
Only once the inclination reaches 78$^{\circ}$ ($\cos\iota = 0.2$) does the
distribution peak at an inclined signal, as in the rightmost plot. However even
for inclinations as great as this, the 90\% credible region cannot exclude
face-on and extends across all orientations from face-on to edge-on. In this
case it is not possible to clearly distinguish the binary orientation. For
values of $\cos\iota < 0.1$ the posterior is peaked at the correct value of
$\iota$ and excludes face-on from the 90\% credible region.

The results shown in Figures \ref{fig:alpha0.1} and \ref{fig:alpha0.5} show the
general features of the distance and inclination distribution.  It is
characterized by three components: one consistent with a face-on signal, one
with an face-off signal and a third contribution peaked around the true values
of distance and inclination.  In all of the cases we have shown, only one or
two of the contributions are significant.  There are, however, cases where we
obtain three distinct peaks in the posterior for the inclination, although
these are rare.  In Appendix B of \cite{Fairhurst:2017mvj}, an approximate
expression for probability associated with each peak was obtained, which is
valid for networks sensitive to a range where a $d_L^2$ prior is still
appropriate.  This provides an analytic expression for the probability
associated to each of the three contributions, as a function of SNR,
inclination, polarization and the network sensitivity to the second
polarization, encoded in the variable $\alpha$.

\begin{table*}[t]
\center
\begin{tabular}{ c || c | c | c || c | c | c || c | c | c || c | c | c }
Network & \multicolumn{3}{c||}{$0^{\circ} \leq \iota < 45^{\circ}$} &
\multicolumn{3}{c||}{$45^{\circ} \leq \iota < 60^{\circ}$} &
\multicolumn{3}{c||}{$60^{\circ} \leq \iota < 75^{\circ}$} &
\multicolumn{3}{c}{$75^{\circ} \leq \iota < 90^{\circ}$ } \\
 & face-on & uncertain & inclined & face-on & uncertain & inclined &
 face-on & uncertain & inclined & face-on & uncertain & inclined   \\
\hline
HL         &   100\% &  0\% & 0\% &  97\% & 3\% & 0\% & 80\% & 18\% & 2\% &  47\% & 32\% & 21\% \\
HLV      &    100\% & 0\%  &0\% &   86\% & 13\% & 1\% & 47\% & 44\% &9\% & 29\% & 27\% & 44\% \\
HLVK    &  100\% & 0\% & 0\% &  78\% & 21\% & 1\% & 27\% & 59\% & 14\% & 17\% & 20\% & 63\% \\
HLVKI   &  100\% & 0\% & 0\% &  67\% & 32\% & 1\% &  7\% & 72\% & 21\% &  7\% & 13\% & 80\% \\
\hline
\end{tabular}
\caption{The table shows the ability of various networks to distinguish the
orientation of a population of binary mergers with given inclination, $\iota$.
For each network and range of $\iota$, we give the percentage of binaries for
which the posterior on the inclination peaks at $\iota = 0$ or $180^\circ$
(face-on) and this peak contains over $90\%$ of the probability; those binaries
for which the recovered inclination peaks at the correct value, and greater
than $90\%$ of the probability is consistent with this peak (inclined); and
those for which the posterior includes significant contributions for both
face-on and inclined orientations (uncertain).  For all networks, essentially
all binaries with $\iota < 45^{\circ}$ will be recovered face-on.  As the
inclination increases further, the ability to clearly identify the binary as
inclined increases significantly with the number of detectors in the network as
this improves the average sensitivity to the second gravitational-wave
polarization.}
\label{table:inclination}
\end{table*}

To get a sense of how accurately binary inclination will be measured, we
simulated a set of $1,000,000$ events uniformly in volume and determined those
which would be observed above the detection threshold of the network (typically
leaving 30,000-80,000 events).  For each event, we then determine whether the
event would be recovered as definitely face-on --- over 90\% of the probability
associated to the face-on (and face-away) components of the distribution ---
definitely inclined or uncertain.  These results are summarized in Table
\ref{table:inclination}, for a series of networks each with an increasing
number of detectors.  For all networks, essentially all events with a true
inclination less than $45^{\circ}$ will be recovered face-on.  Only for those
events with inclination greater than $45^{\circ}$ do we start to be able to
distinguish the orientation.  Between $45$ and $60^{\circ}$, networks with
three or more detectors will classify a small fraction of events as inclined,
and this fraction increases with both the inclination of the system and the
number of detectors (which directly effects the typical value of $\alpha$).
However, even for events which have an inclination greater than $75^{\circ}$,
the LIGO Hanford--Livingston network would recover half as face-on and only
20\% as definitely not.  This improves for the five detector network where less
than 10\% are face-on, and 80\% are clearly identified as being inclined.
We note that similar results have been obtained independently in
 \cite{chen_measuring_2018}.

\begin{figure}[t] \includegraphics[width=\linewidth]{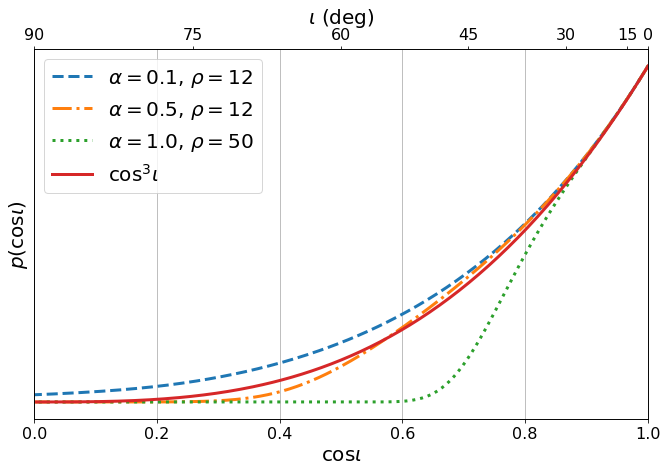}
\caption{The un-normalized marginalized posterior for $\cos \iota$ for a
face-on source as measured for three networks with alignment factors $\alpha =
0.1$, $\alpha = 0.5$, $\alpha = 1.0$ and signal to noise ratio $\rho = 12$,
$\rho = 12$, $\rho = 50$ respectively. The solid line shows the expected
$\cos^3 \iota $ form of the likelihood.}
\label{fig:p_cosi}
\end{figure}

Next, let us consider the general accuracy with which we can measure the
inclination for a binary which is (nearly) face-on.  In this case, the
distribution for the inclination angle can be approximated in a simple way.  If
we begin by assuming that the degeneracy between distance and inclination is
exact, then orientations with $| \cos \iota | \approx 1$ are preferred due to
the prior on the distance.  This can be clearly seen by comparing the second
and third columns of plots in Figure~\ref{fig:alpha0.1}.  The distribution in
the second column (when we don't apply the uniform-in-distance weighting) shows
a broad degeneracy with equal probability along lines of constant $\mathcal{A}
= \cos \iota / d_{L}$.  It is only by applying the distance re-weighting that
the peak shifts more to $\cos \iota = 1$.  For a fixed value of $\iota$, we
wish to integrate over a given distribution, $p(\cos \iota / d_{L})$.  Thus we
obtain
\begin{eqnarray}
p(\cos \iota) &=& \int d_{L}^{2} p(\cos \iota / d_{L}) \text{d} d_{L} \nonumber \\
&=& \int \cos^{3} \iota \mathcal{A}^{-4} p(\mathcal{A}) \nonumber \text{d} \mathcal{A}  \\
&\propto& \cos^{3} \iota
\end{eqnarray}
Thus, it follows that, where the degeneracy holds, the posterior on $\cos\iota$
will be proportional to $\cos^{3}\iota$. In Figure~\ref{fig:p_cosi}, we show
the posterior for three examples of face-on signals : SNR $\rho = 12$ with
$\alpha=0.1$ and $0.5$, and SNR $\rho = 50$ with $\alpha=1$.  All three
distributions follow the $\cos^{3} \iota$ distribution for small inclinations.
The high-SNR signal deviates at around $30^{\circ}$ --- at this inclination
there is enough difference from a circularly polarized signal for larger
inclinations to be disfavoured.  However, for the lower-SNR signals (and also
lower values of $\alpha$) the approximation remains accurate to greater than
$45^{\circ}$.

\begin{figure}[t]
\includegraphics[width=\linewidth]{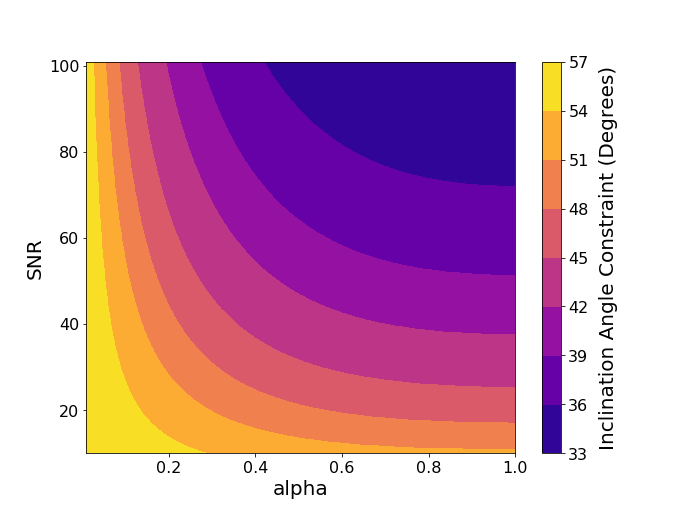}
\caption{This plot shows a detector network's ability to constrain the
inclination of a face-on signal with 90\% confidence. The x-axis shows the
network alignment factor $\alpha$, whereas the y-axis shows the signal-to-noise
ratio (SNR) of the hypothetical gravitational-wave signal. The colour
represents the upper limit on the inclination angle. For weak signals or for
networks which are not very sensitive to the cross polarization, the network
can only constrain the inclination to being less than about 45$^{\circ}$. Even
for the most sensitive detector network detecting the loudest hypothetical
signals, the network would be unable to constrain the inclination to being less
than 30$^{\circ}$. However, we note that at these SNRs, the detector network
may be able to identify higher order modes, which would break the degeneracy
between distance and inclination and allowing us to constrain the inclination
more precisely.}
\label{fig:constraint}
\end{figure}

We can improve the approximation by noting \citep{Fairhurst:2017mvj} that the
SNR lost by projecting an inclined signal onto a circular signal is
\begin{equation}
\Delta \rho^{2} = \frac{\alpha^{2} \rho^{2}}{(1 + \alpha^{2})^{2}} \frac{(1 - \cos \iota)^{4}}{4} \, .
\end{equation}
This loss in SNR leads to a reduction in the likelihood associated with the
inclined signal, which causes the probability distribution to fall off more
rapidly away from $\iota = 1$.  In particular we obtain:
\begin{eqnarray}
p(\cos \iota) &\propto& \cos^{3} \iota \exp\left( - \frac{\Delta \rho^{2}}{2}\right) \, .
\end{eqnarray}
We can use this expression to determine how well a network with sensitivity
$\alpha$ would be able to constrain a signal's inclination $\iota$, given the
SNR of the signal. In Figure~\ref{fig:constraint}, we specifically look at how
tightly we can constrain a face-on signal. We can see that for low-SNR signals
or for networks with little sensitivity to the cross polarization, GW
observations will only be able to constrain the signal to being less than about
45$^{\circ}$. Even with an extremely loud signal and a very sensitive detector
network, we are only able to constrain the signal to about 30$^{\circ}$. It's
important to note here that at these SNRs, higher order modes or precession in
the gravitational-wave signal may be observable. If these are detected, the
degeneracy between distance and inclination would be broken, and we would be
able to more tightly constrain the inclination.

\begin{figure*}[t]
\includegraphics[width=0.99\linewidth]{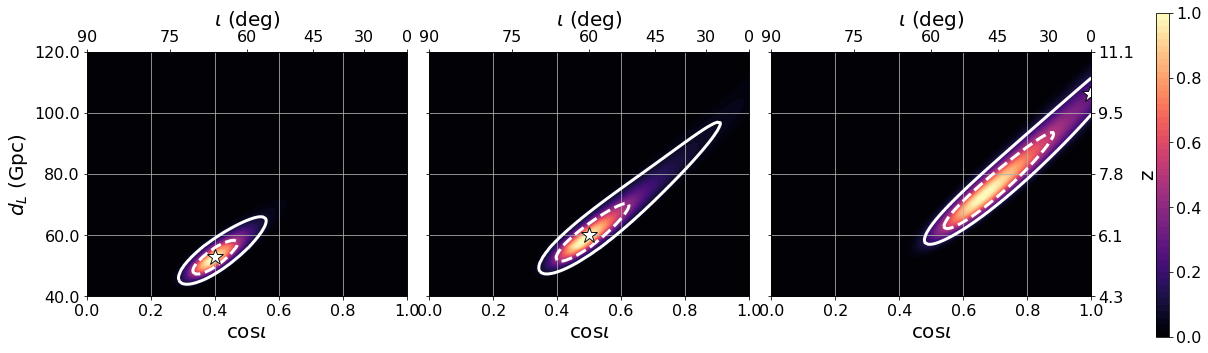}
\caption{Marginalized posterior distribution for a $10M_{\odot}-10 M_{\odot}$
binary black hole at redshift $z = 10$ detected by the Einstein Telescope in
the most sensitive part of the sky, i.e. directly above the detector. Here, the
alignment factor is $\alpha = 1$ and the signal-to-noise ratio is $\rho = 20$.
The white star represents the injected signal at three different inclinations:
$\iota = 66^{\circ}$, $\iota = 60^{\circ}$ and $\iota = 0^{\circ}$. The white
contours represent the 50\% and 90\% confidence intervals obtained from our
simplified model. Note that these contours do not represent the results of full
parameter estimation, as they did in Figure \ref{fig:GW170817}. We use a prior
that is a uniform in comoving volume with a rest frame rate density that
follows the star formation rate \citep{madau_cosmic_2014}. At this redshift the
prior varies by a factor of $\sim 12$ across the degeneracy and now favours
more inclined binaries. Thus binaries that are face-on will be recovered as
being more inclined. The redshift uncertainty $\Delta z / z \sim 40 \%$
dominates the statistical error in the recovery of the binary chirp mass. All
conversions between luminosity distance and redshift assume standard
cosmological parameters \citep{Ade:2015xua}.}
\label{fig:ET-BBH-z10}
\end{figure*}

Finally, it is interesting to consider what effect the inclination distance
degeneracy would have on the mass estimate of binary black holes. GW detectors
actually measure the redshifted mass ${\mathcal {M}_{det}} = (1+z){\mathcal
{M}_{source}}$ where the subscripts denote detector-frame and source-frame
respectively \citep{cutler_gravitational_1994}. There is no way to determine
the redshift directly from the gravitational waveform of a binary black hole.
However the measured value of the luminosity distance can give the redshift if
a cosmology is assumed. In this way, the inclination distance degeneracy will
map to an uncertainty in the rest-frame masses. For the next generation of
gravitational-wave detectors which will be sensitive to BBH mergers throughout
the universe, the uncertainty in
the redshift will likely be the dominant uncertainty in the masses. As such, we
explore the inclination measurement with ET for a BBH merger at a redshift of
$z=10$ with intrinsic masses of a $10M_{\odot}-10 M_{\odot}$ corresponding to a
detector frame chirp mass of ${\mathcal {M}_{det}} = 96 M_{\odot}$. We place
the source directly above the detector, in the most sensitive part of the sky.
In this case, $\alpha = 1$ and $\rho = 20$, where we have assumed standard
cosmology \citep{Ade:2015xua}.

At these cosmological distances, a $d_L^2$ prior for the distance is no longer
appropriate. Rather, we use a distance prior that is uniform in comoving volume
where the rest-frame binary merger rate density follows the cosmic star
formation rate  \citep{madau_cosmic_2014} with a delay between star formation
and binary merger $\Delta t$, and a distribution of delay times $p(\Delta t)
\propto 1/\Delta t$ \citep{totani_cosmological_1997} (see Section 5 of
\cite{mills_localization_2018} for details). The new prior peaks at $z \sim
1.4$. Therefore at $z \sim 10$, the nearer, more inclined binaries are
\textit{a priori} more likely.

In Figure \ref{fig:ET-BBH-z10} we show the marginalized posterior for three
different inclinations: $\iota = 66^{\circ}$, $\iota = 60^{\circ}$ and $\iota =
0^{\circ}$. For the second generation networks in Figure \ref{fig:alpha0.5},
the $\iota = 66^\circ$ ($\cos \iota = 0.4$) source is recovered as face-on.
With the higher signal to noise ratio and improved sensitivity to the second
polarization, ET can identify the signal as edge on. At an inclination of
$\iota = 60^{\circ}$, the degeneracy still extends across $25^{\circ}< \iota <
70^{\circ}$, though smaller inclinations are now excluded from the 90\%
credible interval. This is the effect of the new distance prior which is a
factor of 12 larger at redshift 6 than at redshift 10. Thus, though the 90\%
credible region of the marginalized likelihood extends right up to face-on, the
prior is able to partially break the degeneracy. For less inclined binaries $\iota < 60^{\circ}$, the 90\% probability interval extends up to
face-on.

For the face-on binary in the rightmost plot, the prior shifts the peak of the
posterior away from the true value. Although the value of the likelihood at
face-on and redshift 10 is a factor of 12 larger than it is at an inclination
of $60^{\circ}$ and redshift 6, after the prior re-weighting these two points
in the parameter space are equally likely. If the detector frame chirp mass of
the binary is measured to be ${\mathcal {M}_{det}} = 96 M_{\odot}$, the
degeneracy between the inclination and distance results in ${\mathcal
{M}_{source}} = 96 M_{\odot}$ and ${\mathcal {M}_{source}} = 61 M_{\odot}$
being equally likely.  The detector-frame chirp mass $\mathcal {M}_{det}$ would
be determined to an accuracy similar to the accuracy of the GW phase
measurement $\Delta{\mathcal {M}_{det}}/{\mathcal {M}_{det}} \sim 1/(\rho
{\mathcal {N}_{cycles}})$ \citep{finn_observing_1993, nissanke_exploring_2010}.
Parameter estimation for GW150914 yielded a precision in the detector-frame
mass estimate of $\Delta{\mathcal {M}_{det}}/{\mathcal {M}_{det}} \sim 10\%$
for a comparable SNR
\cite{TheLIGOScientific:2016wfe}.
For a larger mass binary, typically fewer cycles of the waveform will be
visible in the data. However ET's improved sensitivity at low frequencies
compared to LIGO means that we can expect the precision of the detector-frame
mass estimate of GW150914 and the ET binary to be roughly the same. Thus the
broad uncertainty in the intrinsic masses due to the distance inclination
degeneracy $\Delta{\mathcal {M}_{source}}/{\mathcal {M}_{source}} \sim 40\%$
will dominate the total error budget.

\section{Conclusion and Future Work}
Our work demonstrates that even with a network equally sensitive to both
polarizations of the gravitational wave, we would be unable to precisely
measure the inclination or distance of a nearly face-on binary due to a
strong degeneracy between distance and inclination. However, we have focused on
non-spinning binaries and assume that the sky location, masses and arrival
times of the detectors are all known. Introducing these parameters would
increase the uncertainties. Exploring how these parameters affect the overall
measurement of the distance and inclination could give a more accurate summary
a gravitational wave network's ability to measure distance and inclination. 

The degeneracy between inclination and distance described here could be broken
in a few different ways: by using distance or inclination from electromagnetic
measurements, by detecting higher order modes \citep{london_first_2017} and by
measuring precession \citep{vitale_measuring_2018}. Binary neutron star systems
produce a variety of EM signatures, as were observed for GW170817
\citep{Abbott:2017ntl}. Neutron star-black hole binaries (\emph{NSBH}) could
produce EM signatures should the neutron star be tidally disrupted. However,
tidal disruption only happens at relatively small mass ratios
\citep{pannarale_aligned_2015}. For larger mass ratios, the neutron star
plunges into the black hole creating a deformity which rings down.
Interestingly, both precession and higher modes have a larger effect on the
gravitational waveform at higher mass ratio \citep{kidder_coalescing_1995,
varma_gravitational-wave_2014}. The polarizations of the higher modes have a
different dependence on the inclination, and the precession of the orbital
plane would result in changing amplitudes for the plus and cross polarizations.
These effects can make it easier to identify the inclination angle
\citep{london_first_2017, varma_gravitational-wave_2014, Graff:2015bba,
Bustillo:2015qty, vitale_measuring_2018}. For NSBH, the degeneracy can thus be broken by either
information from the EM emission or from higher modes or precession.
\cite{vitale_measuring_2018} demonstrated that precession would break the
distance inclination degeneracy in NSBH for a few binaries with a few values of
the precession angle and large, highly spinning black holes. It would be an
interesting follow up to this study to explore this with a realistic
distribution of spins, to see when precession plays a significant role in
measuring binary parameters.


\bibliographystyle{aasjournal}
\bibliography{inclination_paper}
\end{document}